\documentclass[a4paper,english,cleveref, autoref,thm-restate,nolineno]{socg-lipics-v2021}

\graphicspath{{./figs/}}

\hideLIPIcs
\usepackage{tikz-cd}
\usepackage{tikz}

\newcommand{\R}{
	\mathbb{R}
}

\DeclareMathOperator*{\argmin}{arg\,min}

\newcommand {\mm}[1] {\ifmmode{#1}\else{\mbox{\(#1\)}}\fi}
\newcommand{\Rspace}        {\mm{\mathbb{R}}}
\newcommand{\Xspace}        {\mm{{X}}}
\newcommand{\Yspace}        {\mm{{Y}}}

\newcommand{\etal}{{et al.}}

\title{Measure-Theoretic Reeb Graphs and Reeb Spaces}

\titlerunning{Measure-Theoretic Reeb Graphs} 

\author{Qingsong Wang}{University of Utah, Salt Lake City, USA
}{qswang92@gmail.com}{https://orcid.org/0000-0002-5106-2637}{}

\author{Guanqun Ma}{University of Utah, Salt Lake City, USA
}{guanqun.ma@utah.edu}{https://orcid.org/0000-0001-8102-3172}{}

\author{Raghavendra Sridharamurthy}{University of Utah, Salt Lake City, USA
}{g.s.raghavendra@gmail.com}{https://orcid.org/0000-0001-8463-0488}{}

\author{Bei Wang}{University of Utah, Salt Lake City, USA
}{beiwang@sci.utah.edu}{https://orcid.org/0000-0002-9240-0700}{}

\authorrunning{Q. Wang, G. Ma, R.  Sridharamurthy, and B. Wang} 

\Copyright{Qingsong Wang, Guanqun Ma, Raghavendra Sridharamurthy, and Bei Wang}



\category{} 

\funding{This project was partially supported by NSF DMS 2301361, NSF IIS 2145499, NSF IIS 1910733, DOE DE SC0021015, and DOE DE SC0023157.}

\acknowledgements{}

\EventEditors{Wolfgang Mulzer and Jeff M. Phillips}
\EventNoEds{2}
\EventLongTitle{40th International Symposium on Computational Geometry
(SoCG 2024)}
\EventShortTitle{SoCG 2024}
\EventAcronym{SoCG}
\EventYear{2024}
\EventDate{June 11-14, 2024}
\EventLocation{Athens, Greece}
\EventLogo{socg-logo}
\SeriesVolume{293}
\ArticleNo{XX}     

\begin{document}

\maketitle


\begin{abstract}
    A Reeb graph is a graphical representation of a scalar function on a topological space that encodes the topology of the level sets.
    A Reeb space is a generalization of the Reeb graph to a multiparameter function.
    In this paper, we propose novel constructions of Reeb graphs and Reeb spaces that incorporate the use of a measure.
    Specifically, we introduce measure-theoretic Reeb graphs and Reeb spaces when the domain or the range is modeled as a metric measure space (i.e.,~a metric space equipped with a measure).
    Our main goal is to enhance the robustness of the Reeb graph and Reeb space in representing the topological features of a scalar field while accounting for the distribution of the measure. 
    We first introduce a Reeb graph with local smoothing and prove its stability with respect to the interleaving distance. 
    We then prove the stability of a Reeb graph of a metric measure space with respect to the measure, defined using the distance to a measure or the kernel distance to a measure, respectively.
\end{abstract}

\section{Introduction}
\label{sec:introduction}

A Reeb graph~\cite{Reeb1946} is a topological descriptor that captures the evolution of level sets of a scalar function.
Specifically, given $f: \Xspace \to \Rspace$ defined on a topological space $\Xspace$ with enough regularity, the Reeb graph of $f$ is a  graph where each node corresponds to a critical point of $f$ and each edge captures the relationships among the connected components of the level sets of $f$.
A Reeb space is a generalization of the Reeb graph to a multiparameter function $f: \Xspace \to \Rspace^d$. 
Reeb graphs and Reeb spaces are popular in topological data analysis and visualization; see~\cite{BiasottiGiorgiSpagnuolo2008, HeineLeitteHlawitschka2016, YanMasoodSridharamurthy2021} for surveys.

In this paper, we introduce measure-theoretic Reeb graphs, extensions to the conventional Reeb graph constructions that integrate metric measure spaces---metric spaces endowed with probability measures---to enhance their robustness in capturing the topological features. 
We argue that a metric measure space arises naturally in data.
In many data science applications, we would like to associate weights to data points in the domain or function values in the range, which represent how much we trust these data points or how important their corresponding features are. 
Conventional Reeb graphs, however, do not take into consideration the data distributions and (possibly) non-uniform importance of data points, leading to discrepancies between the represented and actual topologies of the data. 
For example, a significant loop in the Reeb graph might be caused by a sparse set of data points or lie in regions of low importance in function values. 
Our measure-theoretic approach allows Reeb graphs to capture robust topology in data, in line with recent advances in topological data analysis for building robust topological descriptors~\cite{blumberg2014robust,PhillipsWangZheng2015,BuchetChazalOudot2016}.
     Our contributions include:
\begin{itemize}
\item We define a Reeb graph of a metric measure space where the domain is equipped with a measure, and present two stability results:
\begin{itemize}
\item We first introduce a Reeb graph with local smoothing (\cref{def:local-smoothed-Reeb-graph}) and prove its stability with respect to the interleaving distance (\cref{lem:stability-local-smooth}); 
\item We then prove the stability of a Reeb graph of a metric measure space with respect to the measure, defined using the distance to a measure~\cite{BuchetChazalOudot2016} and the kernel distance to a measure~\cite{PhillipsWangZheng2015}, respectively (\cref{thm:stability-dtm-smoothed} and \cref{thm:stability-dmk-smoothed}). 
\end{itemize} 
\item We expand the measure-theoretic construction to consider a measure on the range, referred to as a range-integrated Reeb graph (\cref{def:reeb-graph-respects-measure}), and prove its stability (\cref{prop:reeb-graph-range-measure-stability}).
\item We extend our measure-theoretic constructions (\cref{def:local_smoothed_Reeb_space} and  \cref{def:reeb-space-respects-measure}) and stability results to Reeb spaces (\cref{thm:stability-dtm-smoothed-reeb-space},  \cref{thm:stability-dmk-smoothed-reeb-space}, and \cref{prop:reeb-space-range-measure-stability}).
\item We define a geometric notion of interleaving distance between Reeb spaces (\cref{def:reeb_space_interleaving_distance}) that generalizes that of Reeb graphs and prove the stability of Reeb spaces with respect to this interleaving distance (\cref{thm:reeb_space_stability}).
\end{itemize}

\section{Related Work}
\label{sec:related-work}
\subparagraph*{Reeb graphs and Reeb spaces.}~A Reeb graph~\cite{Reeb1946} is a topological abstraction of the level sets of a scalar function.
A Reeb space~\cite{EdelsbrunnerHarerPatel2008} is analogous to Reeb graphs for a multiparameter function.
Theoretical investigations of Reeb graphs, Reeb spaces, and their variants (in particular, Mapper constructions~\cite{NicolauLevineCarlsson2011})   have been quite active, exploring their distances, information content~\cite{DeyMemoliWang2017, CarriereOudot2018}, stability~\cite{DeSilvaMunchPatel2016,BauerGeWang2014,memoli2018metric,BauerLandiMemoli2020,BauerMunchWang2015,BakkeBjerkevik2021,BotnanLesnick2018, BauerLandiMemoli2020, CarriereOudot2018}, and convergence~\cite{Babu2013, MunchWang2016, DeyMemoliWang2017, CarriereMichelOudot2018,BrownBobrowskiMunch2021}.

There are a number of distances proposed for Reeb graphs and their variants, such as interleaving distance~\cite{BubenikDeSilvaScott2017,ChazalCohenSteinerGlisse2009,DeSilvaMunchPatel2016,MorozovBeketayevWeber2013,MunchStefanou2019,ChambersMunchOphelders2021}, functional distortion distance~\cite{BauerGeWang2014, BauerMunchWang2015}, functional contortion distance~\cite{BauerBjerkevikFluhr2022}, edit distance~\cite{DiFabioLandi2016,BauerFabioLandi2016, BauerLandiMemoli2020, SridharamurthyMasoodKamakshidasan2020},  Gromov-Hausdorff distance~\cite{CarriereOudot2017,TouliWang2019}, and bottleneck distance~\cite{CarriereOudot2017}; see~\cite{BollenChambersLevine2021,YanMasoodSridharamurthy2021} for surveys.
In particular, de Silva \etal~\cite{DeSilvaMunchPatel2016} introduced an interleaving distance that quantifies the similarity between Reeb graphs by utilizing a smoothing construction.
The smoothing idea was further expanded by Munch and Wang~\cite{MunchWang2016}, where they proved the convergence between the Reeb space and Mapper~\cite{SinghMemoliCarlsson2007} in terms of the interleaving distance between their corresponding categorical representations.
Bauer \etal~\cite{BauerMunchWang2015} showed that the interleaving distance is strongly equivalent to the functional distortion distance~\cite{BauerGeWang2014}.  
In this paper, we introduce a local smoothing idea and define an interleaving distance between Reeb spaces that generalizes that of Reeb graphs and prove the stability of Reeb spaces with respect to this interleaving distance.  

Reeb graphs and their variants have been widely used in data analysis and visualization, including shape analysis~\cite{HilagaShinagawaKohmura2001,TungSchmitt2005,TiernyVandeborreDaoudi2009, DeyFanWang2013},  flexible isosurfaces~\cite{CarrSnoeyinkVanDePanne2010}, isosurface denoising~\cite{WoodHoppeDesbrun2004}, data skeletonization~\cite{GeSafaBelkin2011}, topological quadrangulations~\cite{HetroyAttali2003}, loop surgery~\cite{TiernyGyulassySimon2009}, feature tracking~\cite{ChenObermaierHagen2013}, and metric reconstruction of filament structures~\cite{ChazalSun2014}. 
See~\cite{BiasottiGiorgiSpagnuolo2008, YanMasoodSridharamurthy2021} for more applications in computer graphics and data visualization, respectively. 

\subparagraph*{Metric measure spaces.}
A metric measure space is a metric space equipped with a probability measure, providing a natural framework for statistical inference, machine learning, and data analysis~\cite{Sturm2006}. This concept is particularly relevant in real-world data, often sampled from probabilistic distributions, with inherent distance relationships among data points.
In machine learning, metric measure spaces have been used in the study of generative models~\cite{arjovsky2017wasserstein}, graph learning~\cite{chen2022weisfeiler}, and natural language processing~\cite{alvarez2018gromov}.
In topological data analysis, metric measure spaces are instrumental in developing statistically robust persistent homology invariants~\cite{blumberg2014robust,BuchetChazalOudot2016}, studying functional data~\cite{hang2019topological} and providing measure-theoretic perspective on Vietoris-Rips complexes~\cite{adamaszek2018metric,adams2021persistent}.

\subparagraph*{Robust geometric inferences.}
Chazal \etal~\cite{ChazalCohen-SteinerMerigot2011,BuchetChazalOudot2016} introduced the distance to a measure function that supports geometric inferences that are robust to noise and outliers.
As an alternative method, Phillips \etal~\cite{PhillipsWangZheng2015} showed that robust geometric inference of a point cloud can be achieved by examining its kernel density estimate, and subsequently, the kernel distance.
The kernel distance enjoys similar reconstruction properties of distance to a measure, and additionally possesses small coresets~\cite{Phillips2013} for inference tasks.
These robust techniques enhance the resilience of geometric inference against noise and outliers, and are utilized in this paper to attune the measures on metric measure spaces.

\section{Background on Reeb Graphs and Reeb Spaces}
\label{sec:background}

A Reeb graph~\cite{Reeb1946} starts with a topological space $\Xspace$ equipped with a continuous real-valued function $f: \Xspace \to \Rspace$.
It captures the evolution of the level sets of $f$.  
Unless otherwise specified, we always work with continuous functions in this paper. 

\begin{definition}[Reeb graph]
The Reeb graph is the quotient space $R(\Xspace, f) := \Xspace/{\sim_f}$ obtained by identifying equivalent points where, for every $x, y \in \Xspace, x \sim_f y$ if and only if $x$ and $y$ belong to the same connected component of the level set $f^{-1}(f(x))$.
\end{definition}

\begin{figure}[!ht]
    \centering
    \includegraphics[width=0.4\columnwidth]{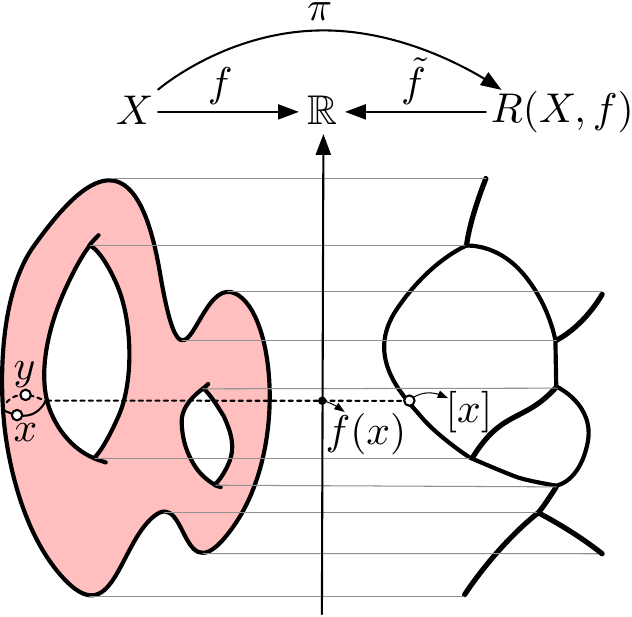}
    \vspace{-2mm}
    \caption{An example of a Reeb graph.} 
    \label{fig:Reeb-graph}
     \vspace{-4mm}
\end{figure}

By construction, as shown in \cref{fig:Reeb-graph}, there is a natural quotient map $\pi: \Xspace \to R(\Xspace, f)$ that sends a point $x\in \Xspace$ to its equivalence class $[x] \in R(\Xspace, f)$.
Meanwhile, $f$ naturally induces a function $\tilde{f}: R(\Xspace, f) \to \R$ defined as $\tilde{f}([x]) = f(x)$.
With some appropriate regularity conditions (for example, $f$ being a piecewise linear function on a finite simplicial complex or a Morse function on a compact manifold), the Reeb graph $R(\Xspace, f)$ is a finite graph and $\tilde{f}$ is a monotonic function on the edges of $R(\Xspace, f)$.
The pair $(\Xspace, f)$ is referred to as an \emph{$\Rspace$-space}~\cite{DeSilvaMunchPatel2016}. 
In this paper, we assume that $\Xspace$ and $f$ are regular enough (e.g. constructible $\Rspace$-spaces~\cite{DeSilvaMunchPatel2016}) so that the Reeb graph $R(\Xspace, f)$ is a finite graph.
We will use this regularity assumption of Reeb graphs throughout the paper.

Let $(\Xspace, f)$ and $(\Yspace, g)$ be two $\Rspace$-spaces.
Following the terminology in~\cite{ChambersMunchOphelders2021}, we say that a continuous map $\phi: \Xspace \to \Yspace$ is a \emph{function preserving map} if $f = g \circ \phi$. 
A function preserving map $\phi:X\to Y$ induces a map $\tilde{\phi}: R(\Xspace, f) \to R(\Yspace, g)$ between the Reeb graphs by sending $[x]$ to $[\phi(x)]$.
Additionally, $\tilde{\phi}$ is also a function preserving map between $(R(\Xspace, f), \tilde{f})$ and $(R(\Yspace, g), \tilde{g})$.
This comes from the universal property of quotient maps; for a proof in the setting of Reeb graphs, see~\cite[Proposition~2.8]{DeSilvaMunchPatel2016}.

To simplify the notation, we write a Reeb graph $R(\Xspace, f)$ as $G:=(G, f)$ with $G$ being a finite graph and $f$ being a real-valued function on $G$ such that $f$ is monotonic on each edge of $G$. We omit $f$ from $(G,f)$ when it is clear from the context. In particular, $G$ is a special case of an $\R$-space.
We say two Reeb graphs are isomorphic if there exist function preserving maps between them that are inverse to each other.

We review the smoothing of Reeb graphs \cite{DeSilvaMunchPatel2016} that facilitates the study of the stability of Reeb graphs.
It is used to define the interleaving distance between Reeb graphs.

\begin{definition}[Smoothing of Reeb graph~\cite{DeSilvaMunchPatel2016}]\label{def:smoothing-Reeb-graph}
	Given a Reeb graph $G$, the $\varepsilon$-smoothing of $G$ is defined as the Reeb graph of the function:
	\begin{equation*}
		\begin{array}{r@{\hspace{0.2cm}}l}
			f_{\varepsilon}:\quad G \times[-\varepsilon, \varepsilon] & \longrightarrow \mathbb{R} \\
			(x, t)                                                              & \longmapsto f(x)+t .
		\end{array}
	\end{equation*}
That is, the \emph{$\varepsilon$-smoothing of a Reeb graph} is the quotient space $G \times[-\varepsilon, \varepsilon] / \sim_{f_{\varepsilon}}$, denoted as $S_\varepsilon(G, f)$.
\end{definition}

\begin{figure}[!ht]
    \centering
    \includegraphics[width=0.5\columnwidth]{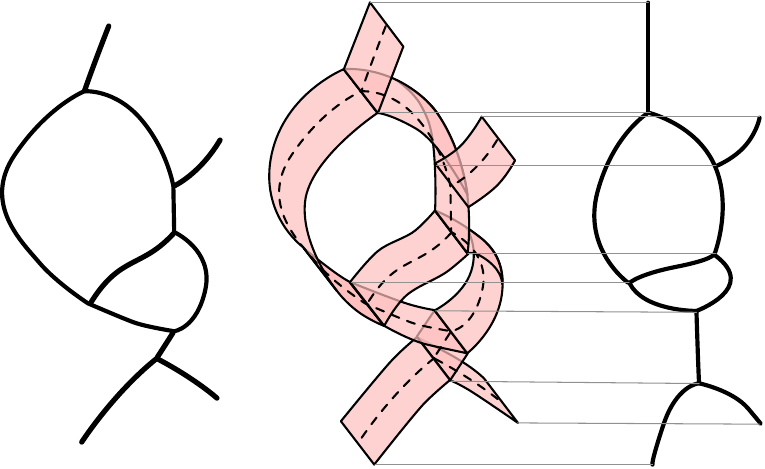}
    \vspace{-2mm}
    \caption{From left to right: a Reeb graph $G$, its $\varepsilon$-thickening with a function $f_{\varepsilon}$, and the Reeb graph of the $\varepsilon$-thickening.}
    \label{fig:eps-smoothing}
\end{figure} 
\noindent The space $G \times[-\varepsilon, \varepsilon]$ is referred to as the \emph{$\varepsilon$-thickening} of $G$. Then the $\varepsilon$-smoothing is the Reeb graph of the $\varepsilon$-thickening. See \cref{fig:eps-smoothing} for an example, where the $\varepsilon$-thickening is tilted slightly to reveal its structure. 
We have the following maps associated with the smoothing of a Reeb graph:
\begin{itemize}
	\item The zero-section inclusion $\eta: G \to S_\varepsilon(G, f)$ is defined as $\eta(x) = [x, 0]$, where we use $[x, 0]$ to denote the equivalence class of $(x, 0)$ in $S_\varepsilon(G, f)$;
	\item Let $\phi: (G, f) \to (H, h)$ be a function preserving map between two Reeb graphs. Then we have the induced map $S_\varepsilon[\phi]$ between their smoothings $S_\varepsilon[\phi]: S_\varepsilon(G, f) \to S_\varepsilon(H, h)$ defined as $\phi_\varepsilon([x, t]) = [\phi(x), t]$.
\end{itemize}
\noindent With the above preparations, we can now present the definition of interleaving distance between Reeb graphs introduced by de Silva et al.~\cite{DeSilvaMunchPatel2016}.
\begin{definition}[Interleaving distance {\cite[Definition~4.35]{DeSilvaMunchPatel2016}}]\label{def:interleaving}
	For any $\epsilon>0$, an $\varepsilon$-interleaving between two Reeb graphs $(G, f)$ and $(H, h)$
	is a pair of maps, $\phi:(G, f) \to S_{\varepsilon}(H, h)$ and $\psi:(H, h) \to$ $S_{\varepsilon}(G, f)$ such that the diagram
	\[\begin{tikzcd}
			{(G, f)} && {S_\varepsilon(G, f)} && {S_{2\varepsilon}(G, f)} \\
			\\
			{(H, h)} && {S_\varepsilon(H, h)} && {S_{2\varepsilon}(H, h)}
			\arrow[from=1-1, to=1-3]
			\arrow[from=1-3, to=1-5]
			\arrow["{S_\varepsilon[\phi]}"'{pos=0.1}, color={rgb,255:red,92;green,92;blue,214}, dashed, from=1-3, to=3-5]
			\arrow[from=3-3, to=3-5]
			\arrow[from=3-1, to=3-3]
			\arrow["\phi"'{pos=0.1}, color={rgb,255:red,92;green,92;blue,214}, from=1-1, to=3-3]
			\arrow["\psi"{pos=0.1}, color={rgb,255:red,222;green,70;blue,23}, from=3-1, to=1-3]
			\arrow["{S_\varepsilon[\psi]}"{pos=0.1}, color={rgb,255:red,222;green,70;blue,23}, dashed, from=3-3, to=1-5]
		\end{tikzcd}\]
	commutes,
	where $S_\varepsilon[\phi]$ is the map induced by $\phi: G \times [-\varepsilon, \varepsilon] \to S_\varepsilon(H, h) \times [-\varepsilon, \varepsilon]$ defined as $\phi(x, t) = (\phi(x), t)$.
	The interleaving distance $d_I((G, f),(H, h))$ is defined as
	\begin{equation*}
		d_I((G, f),(H, h))=\inf _{\varepsilon}\{\text {there exists an } \varepsilon \text {-interleaving of }(G, f) \text { and }(H, h)\}.
	\end{equation*}
\end{definition}

It is shown in \cite{DeSilvaMunchPatel2016} that the interleaving distance is a pseudometric on the set of isomorphism classes of Reeb graphs that takes values in $[0, \infty]$. Additionally, the interleaving distance is zero if and only if the two Reeb graphs are isomorphic.

\begin{proposition}[{\cite[Proposition~4.6]{DeSilvaMunchPatel2016}}]
	Let $(G, f)$ and $(H, h)$ be two Reeb graphs. Then $$d_I((G, f),(H, h)) =0$$ if and only if $(G, f)$ is isomorphic to $(H, h)$.
\end{proposition}

Note that the smoothing can also be applied to the ambient space directly, that is, we consider $S_\varepsilon(\Xspace, f) = \Xspace\times [-\varepsilon, \varepsilon] / \sim_{f_{\varepsilon}}$ where $f_\varepsilon: \Xspace \times [-\varepsilon, \varepsilon] \to \R $ is defined as $f_\varepsilon(x, t) = f(x) + t$. Indeed, the above smoothing construction is discussed in \cite[Definition~4.19]{DeSilvaMunchPatel2016}, and this construction is naturally isomorphic to the one used in~\cref{def:smoothing-Reeb-graph} (in the sense of category theory) as shown in~\cite[Theorem~4.21]{DeSilvaMunchPatel2016}; 
see also~\cref{lem:reeb_space_smoothing}, where we prove this result in the general context of Reeb spaces.
This fact allows the following construction of interleaving maps between Reeb graphs.

\begin{proposition}
\label{prop:interleaving}
	Let $(\Xspace, f)$ and $(\Yspace, g)$ be two $\R$-spaces.
	Then $R(X, f)$ and $R(Y, g)$ are $\epsilon$-interleaved if
	there are function preserving maps $\phi: \Xspace \to \Yspace \times[-\varepsilon, \varepsilon]$ and $\psi: \Yspace \to \Xspace \times[-\varepsilon, \varepsilon]$ such that the following diagram commutes:
	\[\begin{tikzcd}
			{R(\Xspace, f)} && {R(\Xspace\times [-\varepsilon, \varepsilon], f_\varepsilon)} && {R(\Xspace\times [-2\varepsilon, 2\varepsilon], f_{2\varepsilon})} \\
			\\
			{R(\Yspace, g)} && {R(\Yspace\times [-\varepsilon, \varepsilon], g_\varepsilon)} && {R(\Yspace\times [-2\varepsilon, 2\varepsilon], g_{2\varepsilon})}
			\arrow[from=1-1, to=1-3]
			\arrow[from=1-3, to=1-5]
			\arrow["{T_\varepsilon[\tilde{\phi}]}"'{pos=0.1}, color={rgb,255:red,92;green,92;blue,214}, dashed, from=1-3, to=3-5]
			\arrow[from=3-3, to=3-5]
			\arrow[from=3-1, to=3-3]
			\arrow["\tilde{\phi}"'{pos=0.1}, color={rgb,255:red,92;green,92;blue,214}, from=1-1, to=3-3]
			\arrow["\tilde{\psi}"{pos=0.1}, color={rgb,255:red,222;green,70;blue,23}, from=3-1, to=1-3]
			\arrow["{T_\varepsilon[\tilde{\psi}]}"{pos=0.1}, color={rgb,255:red,222;green,70;blue,23}, dashed, from=3-3, to=1-5]
		\end{tikzcd}\]
where $T_\varepsilon[\tilde{\phi}]$ is the map induced by $T_\varepsilon[{\phi}]: \Xspace \times [-\varepsilon, \varepsilon] \to \Yspace \times [-2\varepsilon, 2\varepsilon]$ defined as $$T_\varepsilon[{\phi}](x, t) = (\mathrm{Pr}_1 (\phi(x)), \mathrm{Pr}_2(\phi(x))+ t).$$ We use $\mathrm{Pr}_1$ and $\mathrm{Pr}_2$ to denote the projection maps from $\Yspace \times [-\varepsilon, \varepsilon]$ to $\Yspace$ and $[-\varepsilon, \varepsilon]$ respectively.
\end{proposition}

Finally, we present the following stability result of Reeb graphs $R(\Xspace, f)$ and $R(\Xspace, g)$ that are built from the same ambient space $\Xspace$.
\begin{theorem}[{\cite[Theorem~4.4]{DeSilvaMunchPatel2016}}]\label{thm:stability-reeb}
	Let $R(\Xspace, f)$ and $R(\Xspace, g)$ be two Reeb graphs built from the same ambient space $\Xspace$. Then the interleaving distance defined in~\cref{def:interleaving} satisfies
	\[
		d_I(R(\Xspace, f), R(\Xspace, g)) \leq\|f-g\|_{\infty}.
	\]
\end{theorem}

The Reeb space~\cite{EdelsbrunnerHarerPatel2008} is a natural generalization of the Reeb graph to a multiparameter function $f: \Xspace \to \Rspace^d$.  
Again, we will assume that $X$ and $f$ are regular enough (e.g. they induce a constructible  cosheaf~\cite{Curry2014}).

\begin{definition}[Reeb space]
For any continuous $\R^d$-valued functions $f: \Xspace \to \R^d$, the Reeb space $R(\Xspace, f):= \Xspace/{\sim_f}$ is a quotient space of $\Xspace$ obtained by identifying points that belong to the same connected component of the level set $f^{-1}(c)$ for any $c \in \R^d$.
\end{definition}
As in the case of the Reeb graph, the multiparameter  function $f$ also induces a continuous function $\tilde{f}: R(\Xspace, f) \to \R^d$ on the Reeb space $R(\Xspace, f)$ by $\tilde{f}([x]) = f(x)$ for any $x\in \Xspace$.
For two Reeb spaces $R(\Xspace, f)$ and $R(\Yspace, g)$, a map $\phi:\Xspace \to \Yspace$ is function preserving if $f = g\circ \phi$.
Then the function preserving map $\phi$ induces a map $\tilde{\phi}: R(\Xspace, f) \to R(\Yspace, g)$ on the Reeb spaces by $\tilde{\phi}([x]) = [\phi(x)]$ for any $x\in \Xspace$.
With an abuse of notation, similar to the Reeb graph, we also use the notation $(G, f)$ to denote a Reeb space in \cref{sec:reeb-space}.

\section{Background on Measure-Theoretic Concepts}
\label{sec:measure}

We review measure-theoretic concepts, in particular, the Wasserstein distance between two probability measures on a metric space that originates from optimal transport.
We refer the readers to \cite{villani2009optimal} for more details on the Wasserstein distance.
We also discuss distance to a measure~\cite{ChazalCohen-SteinerMerigot2011,BuchetChazalOudot2016} and kernel distance~\cite{sriperumbudur2010hilbert,PhillipsWangZheng2015} important for robust structural inference.

\begin{definition}[Metric measure space \cite{Sturm2006}]
A \emph{metric measure space} is a triple $(\Xspace, d_\Xspace, \mu)$ where $(\Xspace, d_\Xspace)$ is a metric space and $\mu$ is a probability measure on the Borel $\sigma$-algebra of $\Xspace$.
\end{definition}
Here, we require that the metric
space $(\Xspace, d_\Xspace)$ is complete and separable, and the measure $\mu$ is a locally finite (Borel) probability measure.
For simplicity, we use $\Xspace$ to denote a metric space $(\Xspace, d_\Xspace)$, and $(\Xspace, \mu)$ for a metric measure space,  when $d_\Xspace$ is obvious from the context. 

\begin{definition}[2-Wasserstein distance]
\label{def:Wasserstein_distance}
Let $(\Xspace, d_\Xspace)$ be a metric space and $\mu, \nu$ be two probability measures on $\Xspace$. The $2$-Wasserstein distance between $\mu$ and $\nu$ is defined as
	\begin{equation*}
		W_2(\mu, \nu) = \inf_{\pi \in \Pi(\mu, \nu)} \left(\int_{\Xspace \times \Xspace} d_\Xspace(x, y)^2 d \pi(x, y)\right)^{1 / 2},
	\end{equation*}
	where $\Pi(\mu, \nu)$ is the set of all probability measures on $\Xspace \times \Xspace$ with marginals $\mu$ and $\nu$.
\end{definition}

The distance to a measure function is introduced in \cite{BuchetChazalOudot2016} and it serves as a robust enhancement for geometric inference.
\begin{definition}[Distance to a measure {\cite[Definition~1.1]{BuchetChazalOudot2016}}]
Let $(\Xspace, \mu)$ be a metric measure space and let $m \in  (0,1]$ be a mass parameter. We define the \emph{distance to a measure} function $d_{\mu, m}: \Xspace \to \Rspace$ as
	\begin{equation*}
		d_{\mu, m}: x \in \Xspace \mapsto \sqrt{\frac{1}{m} \int_0^m \delta^2_{\mu, s}(x) d s},
	\end{equation*}
	where $\delta_{\mu, s}$ is defined as	
	$\delta_{\mu, s}: x \in \Xspace \mapsto \inf \{r>0 \mid \mu(\bar{B}(x, r))>s\}$ and
$\bar{B}(x, r)$ denotes the closed ball of radius $r$ centered at $x$.
\end{definition}

The distance to a measure function satisfies the following stability property:
\begin{theorem}[{\cite[Theorem 3.3]{BuchetChazalOudot2016}~for $\R^n$}; {\cite[Proposition 3.14]{buchet2014topological} for general metric spaces}]
	Let $\mu$ and $\nu$ be two probability measures on a metric space $(\Xspace, d_\Xspace)$ and let $m \in (0,1]$ be a mass parameter. Then: 
$\left\|d_{\mu, m}-d_{\nu, m}\right\|_{\infty} \leq \frac{1}{\sqrt{m}} W_2(\mu, \nu),$ where $W_2(\mu, \nu)$ is the 2-Wasserstein distance between $\mu$ and $\nu$.
\end{theorem}

The kernel distance to a measure, as introduced in~\cite{PhillipsWangZheng2015}, also offers an alternative robust enhancement for geometric inference. It is closely related to the kernel density estimation from statistics.
{We generalize this definition from $\R^n$ to general topological spaces by utilizing the notion of \emph{integrally strictly positive definite} kernel functions~\cite{sriperumbudur2010hilbert}.}

\begin{definition}[Integrally strictly positive definite kernel function,~\cite{sriperumbudur2010hilbert}]
\label{def:kernel_function}
	Let $\Xspace$ be topological space. A (Borel) measurable function $K: \Xspace \times \Xspace \to \Rspace$ is called an \emph{integrally strictly positive definite} kernel function if for all finite signed Borel measures $\mu$ on $\Xspace$, there is
	\begin{equation*}
		 \int_{\Xspace \times \Xspace} K(x, x') d \mu(x) d \mu(x') > 0.
	\end{equation*}
\end{definition}

{Examples include the Gaussian kernel function $K(x, x') = \exp(-\|x-x'\|^2/2\sigma^2), \sigma>0$ on $\R^n$, and certain period function $K(x, x') = \exp^{\alpha \cos(x-x')}\cos(\alpha \sin(x-x')), 0 < \alpha \leq 1$ on the circle $\mathbb{S}^1$ (See Section 3.3 of~\cite{sriperumbudur2010hilbert} for details).}
It is shown in~\cite{sriperumbudur2010hilbert} that Defn.~\ref{def:kernel_function} allows us to define a metric on the set of probability measures on $\Xspace$.

\begin{definition}[Kernel distance, {\cite{sriperumbudur2010hilbert,PhillipsWangZheng2015}}]
	\label{def:kernel_distance}
Let $\Xspace$ be a topological space. Let $\mu$ and $\nu$ be two probability measures on $\Xspace$. Let $K$ be an integrally strictly positive definite kernel function. Then the kernel distance $D_K$ between $\mu$ and $\nu$ is defined as
	\begin{equation*}
		D_K(\mu, \nu):= \sqrt{\kappa(\mu, \mu)+\kappa(\nu, \nu)-2 \kappa(\mu, \nu)},
	\end{equation*}
where $\kappa(\mu, \nu)$ is defined as $\kappa(\mu, \nu) := \int_{\Xspace \times \Xspace} K(x, x') d \mu(x) d \nu(x').$
\end{definition}

\begin{theorem}[{\cite[Theorem~7]{sriperumbudur2010hilbert}}]
	Let $\Xspace$ be a topological space. Let $\mu$ and $\nu$ be two probability measures on $\Xspace$. Let $K$ be an integrally strictly positive definite kernel function. Then $D_K$ is a metric on the set of probability measures on $\Xspace$.
\end{theorem}

The kernel distance (Definition~\ref{def:kernel_distance}) is utilized in~\cite{PhillipsWangZheng2015} to define the kernel distance to a measure by considering the kernel distance between a measure and the Dirac delta measure at a point.
We make a slight generalization of the definition to general topological spaces.

\begin{definition}[Kernel distance to a measure, \cite{PhillipsWangZheng2015}]\label{def:kernel-distance}
	Let $\mu$ be a probability measure on a topological space $\Xspace$. Let $K$ be an integrally strictly positive definite kernel function. Then the kernel distance $D_{\mu, K}$ with respect to $\mu$ is a function $D_{\mu, K}: \Xspace \to \R$ defined as
	$
		D_{\mu, K}(x) = D_K(\mu, \delta_x),
	$
	where $\delta_x$ is the Dirac delta measure at $x$.
\end{definition}

Applying the triangle inequality for the kernel distance, we obtain the following stability result of the kernel distance to a measure function.
\begin{theorem}[Stability of kernel distance to a measure, {\cite{PhillipsWangZheng2015}}]\label{thm:stability_kernel_distance}
	Let $\mu$ and $\nu$ be two probability measures on a topological space $\Xspace$. Let $K$ be an integrally strictly positive definite kernel function. Then
$\left\|D_{\mu, K}-D_{\nu, K}\right\|_{\infty} \leq  D_K(\mu, \nu)$,
where $D_K(\mu, \nu)$ is the kernel distance between $\mu$ and $\nu$.
\end{theorem}

\section{Reeb Graphs for Metric Measure Spaces}
\label{sec:reeb-graph}

With the ingredients from \cref{sec:background} and \cref{sec:measure}, we now introduce Reeb graphs for metric measure spaces that are robust to noise in the domain.
We achieve this by utilizing the smoothing operation and using either the distance to a measure~\cite{BuchetChazalOudot2016} or the kernel distance to a measure~\cite{PhillipsWangZheng2015} to define a measure-aware local smoothing factor. 
We first introduce a Reeb graph with local smoothing and prove its stability with respect to the interleaving distance. 
We then prove the stability of Reeb graphs with respect to the measure, defined using the distance to a measure and the kernel distance to a measure, respectively.  

\begin{definition}[Reeb graph with local smoothing]
\label{def:local-smoothed-Reeb-graph}
Let $(\Xspace, f)$ be a $\R$-space.
Let $r: \Xspace \to \Rspace$ be a bounded positive function on $\Xspace$ with $M:=\sup_{x \in \Xspace} r(x)$.
	The function $r$ is viewed as a local smoothing factor.
	Let $\Xspace_r$ denote the space
	$
		\Xspace_r = \{(x, t) \in \Xspace \times [-M, M] \mid |t| \leq r(x) \}.
	$
	Then the function $f$ naturally extends to a function $f_r$ on $\Xspace_r$ by
	$
		f_r(x, t) = f(x) + t.
	$
We define the \emph{$r$-smoothed Reeb graph} of $(\Xspace, f)$ to be the Reeb graph $R(\Xspace_r, f_r)$.
\end{definition}

The standard Reeb graph smoothing is a special case of local smoothing where $r$ is a constant function.
The choice of $r$ can be either the distance to measure function $d_{\mu, m}$ or the kernel distance to a measure function $D_{\mu,K}$. We will call them the distance to a measure smoothed Reeb graph and the kernel distance smoothed Reeb graph,  denoted as $R(\Xspace_{d_{\mu, m}}, f_{d_{\mu, m}})$ and $R(\Xspace_{D_{\mu,K}}, f_{D_{\mu,K}})$ respectively.

We have the following stability result regarding the local smoothing of Reeb graphs.

\begin{lemma}[Stability of locally smoothed Reeb graph]
\label{lem:stability-local-smooth}
Let $\Xspace$ be a topological space and $f$ be a function on $\Xspace$. Additionally, let $r_1$ and $r_2$ be two bounded positive function on $\Xspace$ with $\varepsilon:=\sup_{x \in \Xspace} |r_1(x) - r_2(x)|$. Then the $r_1$-smoothed Reeb graph $R(\Xspace_{r_1}, f_{r_1})$ and the $r_2$-smoothed Reeb graph $R(\Xspace_{r_2}, f_{r_2})$ are $\varepsilon$-interleaved.
\end{lemma}

\begin{proof}
According to~\cref{prop:interleaving}, we need to show the existence of maps $\phi$ and $\psi$ such that the following diagram commutes:
	\[\begin{tikzcd}
			{R(\Xspace_{r_1}, f_{r_1})} && {R(\Xspace_{r_1}\times [-\varepsilon, \varepsilon], f_{r_1, \varepsilon})} && {R(\Xspace_{r_1}\times [-2\varepsilon, 2\varepsilon], f_{r_1, 2\varepsilon})} \\
			&& {} \\
			{R(\Xspace_{r_2}, f_{r_2})} && {R(\Xspace_{r_2}\times [-\varepsilon, \varepsilon], f_{r_2, \varepsilon})} && {R(\Xspace_{r_2}\times [-2\varepsilon, 2\varepsilon], f_{r_2, 2\varepsilon})}
			\arrow["{\eta_{r_1}}", from=1-1, to=1-3]
			\arrow["{\eta_{r_1, \varepsilon}}", from=1-3, to=1-5]
			\arrow["{T_\varepsilon[\phi]}"'{pos=0.1}, color={rgb,255:red,92;green,92;blue,214}, dashed, from=1-3, to=3-5]
			\arrow["{\eta_{r2, \varepsilon}}", from=3-3, to=3-5]
			\arrow["{\eta_{r_2}}", from=3-1, to=3-3]
			\arrow["\phi"'{pos=0.1}, color={rgb,255:red,92;green,92;blue,214}, from=1-1, to=3-3]
			\arrow["\psi"{pos=0.1}, color={rgb,255:red,222;green,70;blue,23}, from=3-1, to=1-3]
			\arrow["{T_\varepsilon[\psi]}"{pos=0.1}, color={rgb,255:red,222;green,70;blue,23}, dashed, from=3-3, to=1-5]
		\end{tikzcd}\]
\noindent In the above diagram, we use the notation $f_{r_1, \varepsilon}$ to denote the function $\Xspace_{r_1} \times [-\varepsilon, \varepsilon] \to \R$ defined as $f_{r_1, \varepsilon}(x, t) = f_{r_1}(x) + t$. We define $f_{r_2, \varepsilon}$, $f_{r_1, 2\varepsilon}$, and $f_{r_2, 2\varepsilon}$ in a similar manner.

Now, let us introduce the maps $\phi$ and $\psi$. We use the pair $(x, t)$ to represent a point in $\Xspace_{r_1}$ and the pair $((x, t), t')$ to represent a point in $\Xspace_{r_1} \times [-\varepsilon, \varepsilon]$ or $\Xspace_{r_1} \times [-2\varepsilon, 2\varepsilon]$.

For any $r>0$, we define the $r$-parameterized projection map $\pi_r: \R \to [-r, r]$ as
\[
\pi_r(t) = \argmin_{-r\leq t'\leq r} |t - t'|.
\]

Recall $r_1$ and $r_2$ are bounded positive functions on $\Xspace$.
We now define the map $\phi: \Xspace_{r_1} \to \Xspace_{r_2} \times [-\varepsilon, \varepsilon]$ as $\phi: (x, t) \mapsto \left((x, \pi_{r_2(x)}(t)), t -\pi_{r_2(x)}(t) \right).$
We want to prove that the map $\phi$ preserves the function value, i.e., $f_{r_1}= f_{r_2, \varepsilon}\circ \phi$ for all $(x, t) \in \Xspace_{r_1}$.
Indeed, for any $(x, t) \in \Xspace_{r_1}$, we have
\[
f_{r_2, \varepsilon}(\phi(x, t)) = f_{r_2}\left(x, \pi_{r_2(x)}(t)\right) + t - \pi_{r_2(x)}(t) = f(x) + t = f_{r_1}(x, t).
\]
We define the map
$\psi: \Xspace_{r_2} \to \Xspace_{r_1} \times [-\varepsilon, \varepsilon]$ as
$\psi: (x, t)  \mapsto \left((x, \pi_{r_1(x)}(t)), t -\pi_{r_1(x)}(t) \right).$
By a similar proof as above, we can show that the map $\psi$ preserves the function value, i.e., $f_{r_2}= f_{r_1, \varepsilon}\circ \psi$ for all $(x, t) \in \Xspace_{r_2}$.
We define $\eta_{r_1}$ to be the inclusion map $\eta_{r_1}: \Xspace_{r_1} \to \Xspace_{r_1}\times [-\varepsilon, \varepsilon]$, that is, $\eta_{r_1}(x, t) = ((x, t), 0)$.
Additionally, let $\eta_{r_1, \varepsilon}$ be the natural inclusion map $\Xspace_{r_1}\times [-\varepsilon, \varepsilon] \to \Xspace_{r_1}\times [-2\varepsilon, 2\varepsilon]$, and the maps $\eta_{r_2}$ and $\eta_{r_2, \varepsilon}$ are defined similarly.
It is straightforward to see that $\eta_{r_1}, \eta_{r_1, \varepsilon}, \eta_{r_2}, \eta_{r_2, \varepsilon}$ are all function preserving maps.
Then we have the following diagram with all the maps preserving function values:
	\[\begin{tikzcd}
			{(\Xspace_{r_1}, f_{r_1})} && {(\Xspace_{r_1}\times [-\varepsilon, \varepsilon], f_{r_1, \varepsilon})} && {(\Xspace_{r_1}\times [-2\varepsilon, 2\varepsilon], f_{r_1, 2\varepsilon})} \\
			&& {} \\
			{(\Xspace_{r_2}, f_{r_2})} && {(\Xspace_{r_2}\times [-\varepsilon, \varepsilon], f_{r_2, \varepsilon})} && {(\Xspace_{r_2}\times [-2\varepsilon, 2\varepsilon], f_{r_2, 2\varepsilon})}
			\arrow["{\eta_{r_1}}", from=1-1, to=1-3]
			\arrow["{\eta_{r_1, \varepsilon}}", from=1-3, to=1-5]
			\arrow["{T_\varepsilon[\phi]}"'{pos=0.1}, color={rgb,255:red,92;green,92;blue,214}, dashed, from=1-3, to=3-5]
			\arrow["{\eta_{r2, \varepsilon}}", from=3-3, to=3-5]
			\arrow["{\eta_{r_2}}", from=3-1, to=3-3]
			\arrow["\phi"'{pos=0.1}, color={rgb,255:red,92;green,92;blue,214}, from=1-1, to=3-3]
			\arrow["\psi"{pos=0.1}, color={rgb,255:red,222;green,70;blue,23}, from=3-1, to=1-3]
			\arrow["{T_\varepsilon[\psi]}"{pos=0.1}, color={rgb,255:red,222;green,70;blue,23}, dashed, from=3-3, to=1-5]
		\end{tikzcd}\]

Since each map preserves function values, we obtain a diagram about maps between Reeb graphs induced by the maps between the spaces in the diagram above.
To conclude the proof, it suffices to show that the induced diagram between Reeb graphs commutes.
We use the notation $\left[\varphi\right]$ to denote the induced map between Reeb graphs for any map $\varphi$ between spaces.
By symmetry, it suffices to show
\begin{itemize}
\item[(i)] $\left[T_\varepsilon[\phi]\right]\circ \left[\eta_{r_1, \varepsilon} \right]=\left[ \eta_{r_2, \varepsilon}\right] \circ \left[\phi\right]$,
		      as maps between $R(\Xspace_{r_1}, f_{r_1})$ and
		      $R(\Xspace_{r_2}\times [-2\varepsilon, 2\varepsilon], f_{r_2, 2\varepsilon})$.
\item[(ii)]  $\left[T_\varepsilon[\psi]\right]\circ \left[\phi\right] = \left[\eta_{r_1, \varepsilon}\right]\circ \left[\eta_{r_1}\right],$
		      as maps between $R(\Xspace_{r_1}, f_{r_1})$ and
		      $R(\Xspace_{r_1}\times [-2\varepsilon, 2\varepsilon], f_{r_1, 2\varepsilon})$.
\end{itemize}
For item (i), let $(x, t) \in \Xspace_{r_1}$, we have
\begin{align*}
		(T_\varepsilon[\phi]\circ \eta_{r_1, \varepsilon})(x, t) & = (T_\varepsilon[\phi])((x, t), 0)
		                                               = (\mathrm{Pr}_1(\phi(x, t)), \mathrm{Pr}_2(\phi(x, t)) + 0) \\
		                                               & = \phi((x, t))                   =  \eta_{r_2, \varepsilon} \circ \phi(x, t), 
	\end{align*}
where $\mathrm{Pr}_1$ and $\mathrm{Pr}_2$ are the projection maps from $\Xspace_{r_1}\times [-\varepsilon, \varepsilon]$ to $\Xspace_{r_1}$ and $[-\varepsilon, \varepsilon]$, respectively.
For item (ii), let $(x, t) \in \Xspace_{r_1}$, we have
	\begin{align*}
		 &(T_\varepsilon[\psi]\circ \phi)(x, t)
		 = (T_\varepsilon[\psi])((x, \pi_{r_2(x)}(t)), t - \pi_{r_2(x)}(t))                                                                                   \\
		                                 & = (\mathrm{Pr}_1(\psi((x, \pi_{r_2(x)}(t)), t - \pi_{r_2(x)}(t))), \mathrm{Pr}_2(\psi((x, \pi_{r_2(x)}(t)), t - \pi_{r_2(x)}(t))) + t - \pi_{r_2(x)}(t)).
	\end{align*}
Note that
	$
		\psi((x, \pi_{r_2(x)}(t))) = ((x, \pi_{r_1(x)}\circ\pi_{r_2}(t)), \pi_{r_2(x)}(t) - \pi_{r_1(x)}\circ\pi_{r_2}(t)).
	$
	Since $|t| \leq r_1(x)$ and $|\pi_{r_2(x)}(t)| \leq t$, $|\pi_{r_2(x)}(t)|\leq r_1(x)$,
	consequently, $\pi_{r_1(x)}\circ\pi_{r_2}(t) = \pi_{r_2(x)}(t)$.
	Therefore, \[
		\psi((x, \pi_{r_2(x)}(t))) = ((x, \pi_{r_2(x)}(t)), 0).
	\]
	Thus, we have
	\begin{align*}
		& (T_\varepsilon[\psi]\circ \phi)(x, t) \\ & = (\mathrm{Pr}_1(\psi((x, \pi_{r_2(x)}(t)), t - \pi_{r_2(x)}(t))), \mathrm{Pr}_2(\psi((x, \pi_{r_2(x)}(t)), t - \pi_{r_2(x)}(t))) + t - \pi_{r_2(x)}(t)) \\
		& = (\mathrm{Pr}_1((x, \pi_{r_2(x)}(t)), 0), \mathrm{Pr}_2((x, \pi_{r_2(x)}(t)), 0) + t - \pi_{r_2(x)}(t)) \\
		& = ((x, \pi_{r_2(x)}(t)), t - \pi_{r_2(x)}(t)) 
	\end{align*}
Note that $\eta_{r_1, \varepsilon}\circ \eta_{r_1}(x, t) = ((x, t), 0)$ and hence the images of $T_\varepsilon[\psi]\circ \phi$ and $\eta_{r_1, \varepsilon}\circ \eta_{r_1}$ are not necessarily the same maps. 
However, when we pass down to the Reeb graph $R(\Xspace_{r_1}\times [-2\varepsilon, 2\varepsilon], f_{r_1, 2\varepsilon})$, the induced maps from $\left(T_\varepsilon[\psi]\circ \phi\right)$ and $\left(\eta_{r_1, \varepsilon}\circ \eta_{r_1}\right)$
agree with each other. Indeed, note that the path $\gamma: [0, 1] \to \Xspace_{r_1}\times [-2\varepsilon, 2\varepsilon]$ defined by
\[
	\gamma: s \mapsto ((x, \pi_{r_2(x)}(t) + s(t - \pi_{r_2(x)}(t))), (1-s)t - \pi_{r_2(x)}(t))
\] 
satisfies $\gamma(0) = ((x, \pi_{r_2(x)}(t)), t - \pi_{r_2(x)}(t)) = (T_\varepsilon[\psi]\circ \phi)(x, t)$ and $\gamma(1) = ((x, t), 0) = (\eta_{r_1, \varepsilon}\circ \eta_{r_1})(x, t)$.
Additionally, $f_{r_1, 2\varepsilon}$ is a constant function on the path $\gamma$ and hence $\left(T_\varepsilon[\psi]\circ \phi\right)$ and $\left(\eta_{r_1, \varepsilon}\circ \eta_{r_1}\right)$ are the same maps from $R(\Xspace_{r_1}, f_{r_1})$ to $R(\Xspace_{r_1}\times [-2\varepsilon, 2\varepsilon], f_{r_1, 2\varepsilon})$. This completes the proof.
\end{proof}

\noindent Now we are ready to prove the stability result of the $d_{\mu, m}$-smoothed Reeb graph and the $D_{\mu,K}$-smoothed Reeb graph with respect to a pair of measures $\mu$ and $\nu$; see Appendix~\ref{sec:reeb-graph-proofs} for proofs using triangle inequalities.

\begin{theorem}[Stability of $d_{\mu, m}$-smoothed Reeb graph]
\label{thm:stability-dtm-smoothed}
	Let $(\Xspace, d_\Xspace, \mu)$ and $(\Xspace, d_\Xspace, \nu)$ be two metric measure spaces and $f, g: \Xspace \rightarrow \mathbb{R}$ be two continuous functions. Let $m \in(0,1]$ be a mass parameter. Then we have
	\begin{equation*}
	d_I(R(\Xspace_{d_{\mu, m}}, f_{d_{\mu, m}}), R(\Xspace_{d_{\nu, m}}, g_{d_{\nu, m}})) \leq  \|f-g\|_{\infty} + \frac{1}{\sqrt{m}}W_2(\mu, \nu).
	\end{equation*}
\end{theorem}

Similarly, for a topological space $\Xspace$ with an integrally strictly positive definite kernel function $K$, we can obtain a similar stability result for the $D_{\mu, K}$-smoothed Reeb graph.
\begin{theorem}[Stability of $D_{\mu,K}$-smoothed Reeb graph]
\label{thm:stability-dmk-smoothed}
Let $\Xspace$ be a topological space. Let $\mu$ and $\nu$ be two probability measures on $\Xspace$.
Let $K$ be an integrally strictly positive definite kernel function on $\Xspace$.
Consider two continuous functions $f, g: \Xspace \rightarrow \mathbb{R}$. Then we have
\[
	d_I(R(\Xspace_{D_{\mu, K}}, f_{D_{\mu, K}}), R(\Xspace_{D_{\nu, K}}, g_{D_{\nu, K}})) \leq  \|f-g\|_{\infty} + D_K(\mu, \nu).
\]
\end{theorem}

\begin{figure*}[!ht]
    \centering
    \includegraphics[width=1.0\columnwidth]{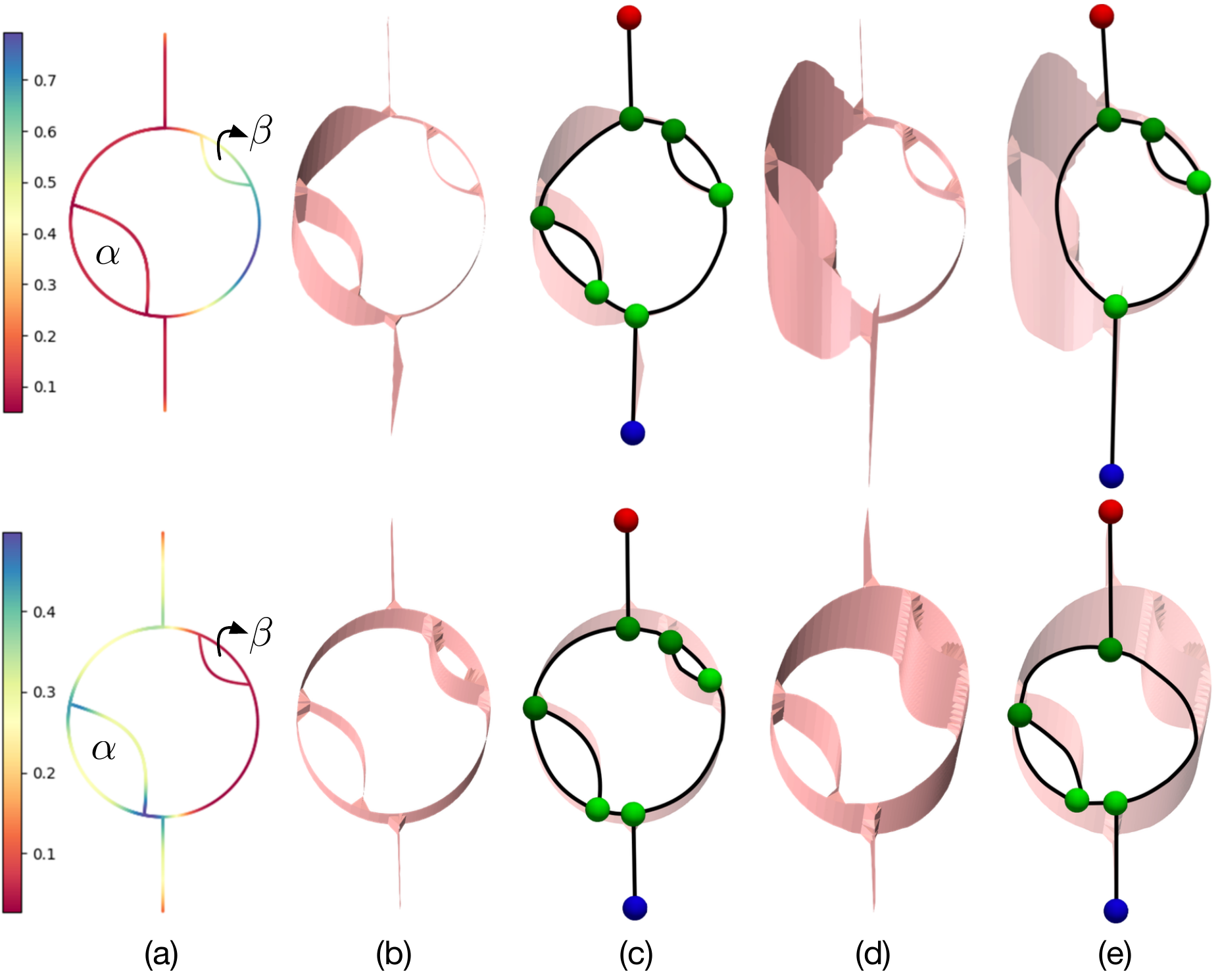}
    \vspace{-4mm}
    \caption{Smoothed Reeb graphs based on distance to a measure  (top) and kernel distance to a measure $D_{\mu, K}$ (bottom) with Gaussian kernel function. From left to right: (a) the original topological space $\Xspace$ colored by a bounded positive function (e.g., $d_{\mu,m}$ or  $D_{\mu, K}$) on
    $\Xspace$; (b) the (locally) thickened spaces with a small $\mu$ value together with (c) the $d_{\mu,m}$-smoothed Reeb graph (top) and the $D_{\mu,K}$-smoothed Reeb graph (bottom); (d)-(e): similar to (b)-(c) with a large $\mu$ value.}
    \label{fig:toy-example-2D}
\end{figure*}

\begin{example}
We use an example in~\cref{fig:toy-example-2D}  to demonstrate the $d_{\mu,m}$-smoothed Reeb graphs (top) and $D_{\mu,K}$-smoothed Reeb graph with Gaussian kernel function (bottom), respectively. 
Our original space $\Xspace$ consists of one large loop containing two small loops $\alpha$ and $\beta$, where $\alpha$ is slightly bigger than $\beta$. 
Since the measure based on $D_{\mu,K}$ considers the larger loop $\alpha$ on the bottom left corner to be more important, the $D_{\mu,K}$-smoothed Reeb graph retains $\alpha$ at a larger $\mu$ value.   
On the other hand, the measure based on $d_{\mu,m}$ emphasizes the significance of the smaller loop $\beta$ on the upper right corner, the $d_{\mu,m}$-smoothed Reeb graph thus retains $\beta$ at a larger $\mu$ value.   
\end{example}

\section{Reeb Spaces for Metric Measure Spaces}
\label{sec:reeb-space}

The stability results in \cref{sec:reeb-graph} extends to the setting of Reeb spaces for metric measure spaces. 
In this section, we use $R(X, f)$ to denote the Reeb space of a multiparameter function $f:\Xspace \to \Rspace^d$. 
We assume that the topological space $X$ and the resulting Reeb space $R(X, f)$ are compact and Hausdorff.
The smoothing and hence the notion of interleaving distance of Reeb graph is extended to Reeb space in~\cite{MunchWang2016} through categorical language.
In this section, we focus on a geometric approach for  smoothing a Reeb space. 
All omitted proofs in this section are given in Appendix~\ref{sec:reeb-space-proofs}.
We first introduce the following notations.
Let $I_\varepsilon := \{t\in \R^d \mid |t|_\infty \leq \varepsilon\}$ be the $\ell_\infty$ ball of radius $\varepsilon$ centered at the origin, it serves as a  higher-dimensional analogue of the 1-dimensional interval. 

\begin{definition}[Smoothing of Reeb space]\label{def:reeb_space_smoothing}
Let $(G, f)$ be a Reeb space.
For any $\varepsilon > 0$, the $\varepsilon$-smoothing $S_\varepsilon(G, f)$ of $(G, f)$ is a Reeb space $R(G\times I_\varepsilon, f_\varepsilon)$ where $f_\varepsilon: G\times I_\varepsilon \to \R^d$ is the continuous function defined by $f_\varepsilon(x, t) = f(x) + t$ for any $(x, t)\in G\times I_\varepsilon$.
\end{definition}

\noindent We now define a geometric notion of interleaving distance between Reeb spaces.

\begin{definition}[Interleaving distance between Reeb spaces]\label{def:reeb_space_interleaving_distance}
	For any $\varepsilon>0$, an $\varepsilon$-interleaving between two Reeb spaces $(G, f)$ and $(H, h)$
	is a pair of maps, $\phi:(G, f) \to S_{\varepsilon}(H, h)$ and $\psi:(H, h) \to$ $S_{\varepsilon}(G, f)$ such that the diagram
	\[\begin{tikzcd}
			{(G, f)} && {S_\varepsilon(G, f)} && {S_{2\varepsilon}(G, f)} \\
			\\
			{(H, h)} && {S_\varepsilon(H, h)} && {S_{2\varepsilon}(H, h)}
			\arrow[from=1-1, to=1-3]
			\arrow[from=1-3, to=1-5]
			\arrow["{S_\varepsilon[\phi]}"'{pos=0.1}, color={rgb,255:red,92;green,92;blue,214}, dashed, from=1-3, to=3-5]
			\arrow[from=3-3, to=3-5]
			\arrow[from=3-1, to=3-3]
			\arrow["\phi"'{pos=0.1}, color={rgb,255:red,92;green,92;blue,214}, from=1-1, to=3-3]
			\arrow["\psi"{pos=0.1}, color={rgb,255:red,222;green,70;blue,23}, from=3-1, to=1-3]
			\arrow["{S_\varepsilon[\psi]}"{pos=0.1}, color={rgb,255:red,222;green,70;blue,23}, dashed, from=3-3, to=1-5]
		\end{tikzcd}\]
	commutes,
	where $S_\varepsilon[\phi]$ is the map induced by $\Phi: G \times I_\varepsilon \to S_\varepsilon(H, h) \times I_\varepsilon$ defined as $\Phi(x, t) = (\phi(x), t)$.
	The interleaving distance $d_I((G, f),(H, h))$ is defined as
	\begin{equation*}
		d_I((G, f),(H, h))=\inf _{\varepsilon}\{\text {there exists an } \varepsilon \text {-interleaving of }(G, f) \text { and }(H, h)\}.
	\end{equation*}
\end{definition}

Suppose the Reeb space $R(\Xspace, f)$ is induced by a continuous function $f: \Xspace \to \R^d$.
Then the $\varepsilon$-smoothing $S_\varepsilon(R(\Xspace, f),\tilde{f})$ is the same as the Reeb space induced by the continuous function $f_\varepsilon: \Xspace \times I_\varepsilon \to \R^d$ on $\Xspace \times I_\varepsilon$ defined by $f_\varepsilon(x, t) = f(x) + t$ for any $(x, t)\in \Xspace \times I_\varepsilon$. Indeed, we have the following lemma.

\begin{lemma}\label{lem:reeb_space_smoothing}
    Let $R(X, f)$ be a Reeb space induced by a continuous function $f: \Xspace \to \R^d$.
	Let $R(\Xspace \times I_\varepsilon, f_\varepsilon)$ be the
   	Reeb space induced by the continuous function $f_\varepsilon: \Xspace \times I_\varepsilon \to \R^d$ on $\Xspace \times I_\varepsilon$ defined by $f_\varepsilon(x, t) = f(x) + t$ for any $(x, t)\in \Xspace \times I_\varepsilon$.
	Then there exists a homeomorphism from $S_\varepsilon(R(\Xspace, f))$ to $R(\Xspace \times I_\varepsilon, f_\varepsilon)$ that preserves function values.
\end{lemma}

As a direct consequence, we have the following extension of Proposition~\ref{prop:interleaving} to Reeb spaces.

\begin{proposition}
    Let $R(\Xspace,f)$ and $R(\Yspace, g)$ be two Reeb spaces induced by continuous functions $f: \Xspace \to \R^d$ and $g: \Yspace \to \R^d$ respectively.
	Then $R(\Xspace, f)$ and $R(\Yspace, g)$ are $\varepsilon$-interleaved if there are function preserving maps $\phi: \Xspace \to \Yspace \times I_\varepsilon$ and $\psi: \Yspace \to \Xspace \times I_\varepsilon$ such that the following diagram commutes:
	\[\begin{tikzcd}
			{R(\Xspace, f)} && {R(\Xspace\times I_\varepsilon, f_\varepsilon)} && {R(\Xspace\times I_{2\varepsilon}, f_{2\varepsilon})} \\
			\\
			{R(\Yspace, g)} && {R(\Yspace\times I_\varepsilon, g_\varepsilon)} && {R(\Yspace\times I_{2\varepsilon}, g_{2\varepsilon})}
			\arrow[from=1-1, to=1-3]
			\arrow[from=1-3, to=1-5]
			\arrow["{T_\varepsilon[\tilde{\phi}]}"'{pos=0.1}, color={rgb,255:red,92;green,92;blue,214}, dashed, from=1-3, to=3-5]
			\arrow[from=3-3, to=3-5]
			\arrow[from=3-1, to=3-3]
			\arrow["\tilde{\phi}"'{pos=0.1}, color={rgb,255:red,92;green,92;blue,214}, from=1-1, to=3-3]
			\arrow["\tilde{\psi}"{pos=0.1}, color={rgb,255:red,222;green,70;blue,23}, from=3-1, to=1-3]
			\arrow["{T_\varepsilon[\tilde{\psi}]}"{pos=0.1}, color={rgb,255:red,222;green,70;blue,23}, dashed, from=3-3, to=1-5]
		\end{tikzcd}\]
where $T_\varepsilon[\tilde{\phi}]$ is the map between Reeb graphs induced by $T_\varepsilon[{\phi}]: \Xspace \times I_\varepsilon \to \Yspace \times I_{2\varepsilon}$ defined as $$T_\varepsilon[\phi](x, t) = (\mathrm{Pr}_1 (\phi(x)), \mathrm{Pr}_2(\phi(x))+ t).$$ Here, we use $\mathrm{Pr}_1$ and $\mathrm{Pr}_2$ to denote the projection maps from $\Yspace \times I_{2\varepsilon}$ to $\Yspace$ and $I_{2\varepsilon}$ respectively.
\end{proposition}

As in the case of Reeb graph, we have the following stability result for Reeb spaces built from the same space with multiparameter functions; see~\cref{sec:reeb-space-proofs} for the proof.

\begin{theorem}\label{thm:reeb_space_stability}
	Let $f, g: \Xspace \to \R^d$ be two bounded continuous functions on $\Xspace$.
	Then the Reeb spaces $R(\Xspace, f)$ and $R(\Xspace, g)$ are $(\|f-g\|_\infty)$-interleaved.
\end{theorem}

With the smoothing of Reeb space, we can also define the Reeb space with local smoothing which in turn allows us to define Reeb spaces for metric measure spaces.

\begin{definition}[Reeb space with local smoothing]
	\label{def:local_smoothed_Reeb_space}
	Let $f: \Xspace \to \R^d$ be a continuous function on $\Xspace$. Additionally, let
		$r: \Xspace \to \Rspace$ be a bounded positive function on $\Xspace$ with $M:=\sup_{x \in \Xspace} r(x)$.
		The function $r$ is viewed as a local smoothing factor.
		We use $\Xspace_r$ to denote the space
		$
			\Xspace_r = \{(x, t) \in \Xspace \times [-M, M]^d \mid |t| \leq r(x) \}.
		$
		Then the function $f$ naturally extends to a function $f_r$ on $\Xspace_r$ by	$f_r(x, t) = f(x) + t.$
	We then defined the \emph{$r$-smoothed Reeb space} of $(\Xspace, f)$ to be the Reeb space $R(\Xspace_r, f_r)$.
\end{definition}

As in the case of Reeb graph, for a metric measure space $(\Xspace, d, \mu)$ with $\R^d$-valued function $f$, we can define the distance to a measure smoothed Reeb graph $R(\Xspace_{d_{\mu, m}}, f_{d_{\mu, m}})$ and the kernel distance smoothed Reeb graph $R(\Xspace_{D_{\mu,K}}, f_{D_{\mu,K}})$ by using $d_{\mu, m}$ and  $D_{\mu,K}$ as the local smoothing factor $r$ in Definition~\ref{def:local_smoothed_Reeb_space}.

By considering the variable $t$ belonging to $I_\varepsilon$ instead of $t\in [-\varepsilon, \varepsilon]$, the exact same proof of~\cref{lem:stability-local-smooth} can be extended to the Reeb space with local smoothing. That is, the Reeb space with local smoothing is stable with respect to the local smoothing factor $r$. Therefore, we have the following stability result for Reeb space with local smoothing. The proof is identical to the proof of~\cref{lem:stability-local-smooth} by simply viewing the parameter $t$ as an element in $\R^d$ instead of $\R$.

\begin{lemma}
	\label{lem:stability-local-smooth-reeb-space}
	Let $(\Xspace, d, \mu)$ be a metric measure space and $f: \Xspace \to \R^d$ be a continuous function.
	Let $r_1, r_2: \Xspace \to \Rspace$ be two bounded positive functions on $\Xspace$ with $\epsilon:=\sup_{x \in \Xspace} |r_1(x) - r_2(x)|$.
	Then the Reeb spaces $R(\Xspace_{r_1}, f_{r_1})$ and $R(\Xspace_{r_2}, f_{r_2})$ are $\epsilon$-interleaved.
\end{lemma}

Likewise, we have the following stability results for Reeb space with local smoothing with respect to functions $d_{\mu, m}$ and $D_{\mu,K}$.

\begin{theorem}
	\label{thm:stability-dtm-smoothed-reeb-space}
	Let $(\Xspace, d_\Xspace, \mu)$ and $(\Xspace, d_\Xspace, \nu)$ be two metric measure spaces. Let $f, g: \Xspace \rightarrow \mathbb{R}^d$ be two continuous functions. Let $m \in(0,1]$ be a mass parameter. Then we have
	\begin{equation*}
	d_I(R(\Xspace_{d_{\mu, m}}, f_{d_{\mu, m}}), R(\Xspace_{d_{\nu, m}}, g_{d_{\nu, m}})) \leq  \|f-g\|_{\infty} + \frac{1}{\sqrt{m}}W_2(\mu, \nu).
	\end{equation*}
\end{theorem}

\begin{theorem}
	\label{thm:stability-dmk-smoothed-reeb-space}
	Let $(\Xspace, d_\Xspace, \mu)$ and $(\Xspace, d_\Xspace, \nu)$ be two metric measure spaces.
Let $K$ be an integrally strictly positive definite kernel function on $\Xspace$.
Let $f, g: \Xspace \rightarrow \mathbb{R}^d$ be two continuous functions. Let $m \in(0,1]$ be a mass parameter.
	Then we have
	\begin{equation*}
	d_I(R(\Xspace_{D_{\mu, K}}, f_{D_{\mu, K}}), R(\Xspace_{D_{\nu, K}}, g_{D_{\nu, K}})) \leq  \|f-g\|_{\infty} + D_K(\mu, \nu).
	\end{equation*}
\end{theorem}


\section{Range-Integrated Reeb Graphs}
\label{sec:range-integrated}

\begin{figure}[!ht]
    \centering
    \includegraphics[width=0.5\linewidth]{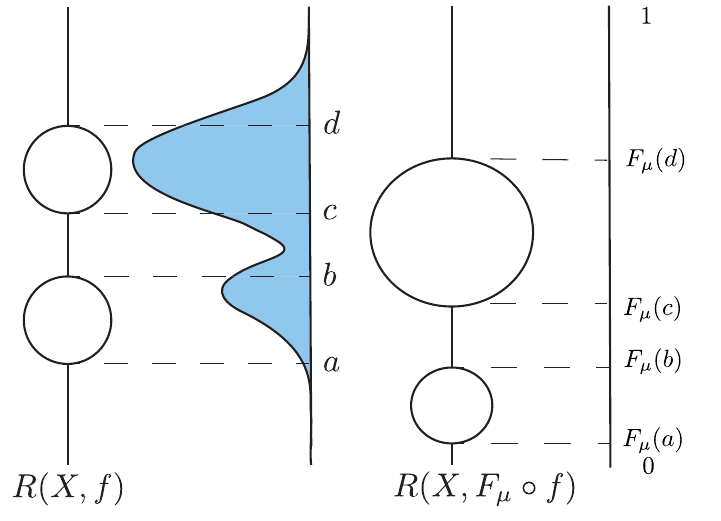}
    \caption{Visualization of a Reeb graph $R(X, f)$ (left) and a range-integrated Reeb Graph $R(X, F_\mu \circ f)$ (right) respectively.}
    \label{fig:range-reeb}
\end{figure}

Our extension of Reeb graphs to metric measure spaces needs not to be limited to measures defined on the domain of the function.
We now extend the Reeb graph construction so that it respects a measure $\mu$ on the range of a function.
For instance, $\mu$ may capture the importance of a feature and we would like to understand  how $\mu$ transforms the shape of a Reeb graph.    
Let $X$ be a topological space and $f: X \to \R$ be a continuous function.
Let $\mu$ be a probability measure on $\R$.
The \emph{cumulative distribution function} (CDF) of $\mu$ is defined as
    \begin{equation*}
        F_{\mu}(x) := \mu((- \infty , x]) = \int_{- \infty}^{x} d \mu.
    \end{equation*}
Therefore, a natural way to adapt the Reeb graph construction when its range comes with a measure $\mu$ is to consider the Reeb graph of the function $F_{\mu} \circ f$. We assume the function $F_{\mu} \circ f$ is regular so that the Reeb graph $R(X, F_{\mu} \circ f)$ is a finite graph.
\begin{definition}[Range-integrated Reeb graph]
\label{def:reeb-graph-respects-measure}
Let $X$ be a topological space and $f: X \to \R$ be a continuous function.
Let $\mu$ be a probability measure on $\R$ whose CDF $F_{\mu}$ is continuous.
Then the \emph{range-integrated Reeb Graph} of $f$ with respect to $\mu$ is defined to be the Reeb graph of $F_{\mu} \circ f$, denoted as $R(X, F_\mu \circ f)$.
\end{definition}
\noindent We provide in~\cref{fig:range-reeb} an example of the Reeb graph that respects a measure on the range of the function.
The intuition behind a range-integrated Reeb graph is that a measure $\mu$ on the range enables the vertical scaling (stretching/shrinking) of a Reeb graph according to $\mu$, which subsequently emphasizes certain topological features according to $\mu$.

In the following, we show that the above construction is stable.
To this end, we utilize the Kolmogorov-Smirnov distance between two probability measures on $\R$.

\begin{definition}[Kolmogorov-Smirnov distance]
\label{def:kolmogorov-smirnov-distance}
Let $\mu$ and $\nu$ be two probability measures on $\R$.
Then the \emph{Kolmogorov-Smirnov (KS) distance} $d_{KS}$ between $\mu$ and $\nu$ is defined as
    \begin{equation*}
        d_{KS}(\mu, \nu) := \sup_{x \in \R} |F_{\mu}(x) - F_{\nu}(x)|.
    \end{equation*}
\end{definition}

\noindent Recall that the Lipschitz constant of a function $f: \R \to \R$ is defined as
   $
        \mathrm{Lip}(f) := \sup_{x, y \in \R} \frac{|f(x) - f(y)|}{|x - y|}.
  $
Then we have the following stability result whose proof, as well as
the proofs of other results in this section are
deferred to Appendix~\ref{sec:range-integrated-proofs}.

\begin{proposition}
\label{prop:reeb-graph-range-measure-stability}
Let $X$ be a topological space and $f, g: X \to \R$ two continuous functions. 
Let $\mu, \nu$ be two probability measures on $\R$ with continuous CDF $F_\mu, F_\nu$ respectively.
Then we have the following inequality:
    \begin{align*}
        d_I(R(X, F_\mu \circ f), R(X, F_\nu \circ g)) &\leq  \min\left\{d_{KS}(\mu, \nu)+ \mathrm{Lip}(F_\mu)||f - g||_{\infty},\right.\\
        &\quad\quad\quad\quad \left. d_{KS}(\mu, \nu) + \mathrm{Lip}(F_\nu)||f - g||_{\infty}, 1 \right\}.
    \end{align*}
\end{proposition}

\noindent Specifically, when $\mu$ approaches $\nu$ and $f$ approaches $g$, the above inequality implies that the interleaving distance between $R(X, F_\mu \circ f)$ and $R(X, F_\nu \circ g)$ approaches to zero.

Let $X$ be a manifold with a Morse function $f$.
Then the nodes of the Reeb graph $R(X, f)$ are the critical points of $f$, i.e., the points $x \in X$ such that the gradient $\nabla f(x) = 0$.
Under some mild conditions, the range-integrated Reeb graph $R(X, F_\mu \circ f)$ rescales the Reeb graph $R(X, f)$ according to the measure $\mu$ on the range of $f$ as in the following proposition.

\begin{proposition}
\label{prop:reeb-graph-range-measure-critical-points}
Let $X$ be a manifold and $f: X \to \R$ be a Morse function.
Let $\mu$ be a probability measure on $\R$.
If the following conditions hold:
\begin{enumerate}
    \item The measure $\mu$ admits a continuously differentiable density function $p_\mu$ with respect to the Lebesgue measure $\lambda$ on $\R$, that is, $\mu(A) = \int_A p_\mu d \lambda$ for any Borel set $A \subset \R$;
    \item The image of $f$ is contained in the interior of the support of $\mu$, that is, for any $x \in X$, $ p_\mu(f(x)) > 0$.
\end{enumerate}
Then the composition $F_\mu \circ f$ is Morse and the
critical points of $F_\mu \circ f$ are the same as the critical points of $f$ with corresponding critical values being $p_\mu(f(x))$ for each critical point $x$ of $f$.
Furthermore, for each critical point $x$ of $f$, the Hessian of $F_\mu \circ f$ at $x$ has the same number of positive and negative eigenvalues as the Hessian of $f$ at $x$.
\end{proposition}

Since the topology of the Reeb graph $R(X, f)$ is determined by the critical points of $f$ and the index of the Hessian of $f$ at each critical point, the above proposition implies that the range-integrated Reeb graph $R(X, F_\mu \circ f)$ maintains the same topology as $R(X, f)$ (under certain conditions) and only stretches/shrinks the Reeb graph $R(X, f)$ according to the measure $\mu$.
In~\cref{fig:range-reeb}, we present a visualization of a comparison between the Reeb graph $R(X, f)$ and the range-integrated Reeb graph $R(X, F_\mu \circ f)$ in the setting of Proposition~\ref{prop:reeb-graph-range-measure-critical-points}.

\section{Range-Integrated Reeb Spaces}
\label{sec:range-reeb-space}

In this section, we extend the range-integrated Reeb Graph construction to Reeb spaces. 
Let \(\mu\) be a probability measure on \(\mathbb{R}^d\). For \(1 \leq i \leq d\), denote by \(\pi_i\) the projection from \(\mathbb{R}^d\) to \(\mathbb{R}\) along the \(i\)-th coordinate, where \(\pi_i(x_1, \dots, x_d) = x_i\). The marginal distribution of \(\mu\) along the \(i\)-th coordinate, \(\mu_i\), is given by \(\mu_i(B) = \mu(\pi_i^{-1}(B))\) for any Borel set \(B \subset \mathbb{R}\).

\begin{definition}[Range-integrated Reeb space]
\label{def:reeb-space-respects-measure}
Let $X$ be a topological space and $f: X \to \R^d$ be a continuous function.
Let $\mu$ be a probability measure on $\R^d$ such that the CDF $F_{\mu_i}$ of $\mu_i$ is continuous for each $1\leq i \leq d$. 
We define the coordinate-wise CDF $F_{\mu}$ of $\mu$ as follows:
$F_{\mu}: \R^d \to \R^d$ as
$
    F_{\mu}(x_1, \dots, x_d) = (F_{\mu_1}(x_1), \dots, F_{\mu_d}(x_d))
$, where $F_{\mu_i}$ is the CDF of $\mu_i$. 
Then, the \emph{range-integrated Reeb space} of $f$ with respect to $\mu$ is defined to be the Reeb space of $F_{\mu} \circ f$, denoted as $R(X, F_\mu \circ f)$.
\end{definition}

Following the same intuition as in the case of range-integrated Reeb graphs, the above construction enables stretching/shrinking of a Reeb space according to a measure $\mu$ on the range of a function $f$.
We will show the stability of the range-integrated Reeb space in the following proposition whose proof is deferred to Appendix~\ref{sec:range-reeb-space-proofs}.

\begin{proposition}
\label{prop:reeb-space-range-measure-stability}
Let $X$ be a topological space and $f, g: X \to \R^d$ two continuous functions.
Let $\mu, \nu$ be two probability measures on $\R^d$ such that their coordinate-wise CDFs $F_\mu, F_\nu$ are continuous.
Then we have the following inequality:
    \begin{align*}
        d_I(R(X, F_\mu \circ f), R(X, F_\nu \circ g)) &\leq  \min\left\{ \mathrm{Lip}(F_\mu)||f - g||_{\infty} + \max_{1\leq i\leq d} \{
            d_{KS}(\mu_i, \nu_i) \},\right.\\
        &\quad\quad\quad\quad \left. \mathrm{Lip}(F_\nu)||f - g||_{\infty} + \max_{1\leq i\leq d} \{
            d_{KS}(\mu_i, \nu_i)
        \}, 1 \right\}.
    \end{align*}
where the Lipschitz constant of a vector valued fucntion $f: \R^d \to \R^d$ is defined with respect to the $\ell_\infty$ norm, that is,
$\mathrm{Lip}(f) := \sup_{x, y \in \R^d} \frac{\|f(x) - f(y)\|_{\infty}}{ \|x - y\|_{\infty}}.$
\end{proposition}

\section{Conclusion and Discussion}
\label{sec:conclusion}

In this work, we present a novel theoretical framework for Reeb graphs and Reeb spaces, utilizing metric measure spaces in either the domain or the range. Our findings demonstrate the stability of both Reeb graph and Reeb space constructions against perturbations of the function and the measure, thereby offering robust improvements for these topological descriptors. 
Additionally, as one key component of our framework, we define a geometric notion of interleaving distance between Reeb spaces that generalizes that of Reeb graphs and prove the stability of Reeb spaces with respect to this interleaving distance.
Moving forward, we will explore the utility of our framework in topological data analysis and visualization. 
We will also study the stability of Reeb graphs using distances between their level set persistence diagrams, again in the context of metric measure spaces.

\bibliography{refs-reeb}

 \appendix
 \section{Detailed Proofs}

 \subsection{Detailed Proofs for Section~\ref{sec:reeb-graph}}
 \label{sec:reeb-graph-proofs}
 In this section, we provide detailed proofs for the results in~\cref{sec:reeb-graph}.

 \begin{proof}[Proof of Theorem~\ref{thm:stability-dtm-smoothed}]
 	By the triangle inequality of interleaving distance, we have
 	\begin{align*}
 	d_I(R(\Xspace_{d_{\mu, m}}, f_{d_{\mu, m}}), R(\Xspace_{d_{\nu, m}}, g_{d_{\nu, m}})) & \leq d_I(R(\Xspace_{d_{\mu, m}}, f_{d_{\mu, m}}), R(\Xspace_{d_{\mu, m}}, g_{d_{\mu, m}})) \\
 	& + d_I(R(\Xspace_{d_{\mu, m}}, g_{d_{\mu, m}}), R(\Xspace_{d_{\nu, m}}, g_{d_{\nu, m}}))\\
 	& \leq \|f-g\|_{\infty} + \frac{1}{\sqrt{m}}W_2(\mu, \nu),
 	\end{align*}
 	where the last step uses~\cref{thm:stability-reeb} and~\cref{lem:stability-local-smooth}.
 \end{proof}

 \begin{proof}[Proof of Theorem~\ref{thm:stability-dmk-smoothed}]
 		By the triangle inequality of interleaving distance, we have
 		\begin{align*}
 		d_I(R(\Xspace_{D_{\mu, K}}, f_{D_{\mu, K}}), R(\Xspace_{D_{\nu, K}}, g_{D_{\nu, K}})) & \leq d_I(R(\Xspace_{D_{\mu, K}}, f_{D_{\mu, K}}), R(\Xspace_{D_{\mu, K}}, g_{D_{\mu, K}})) \\
 		& + d_I(R(\Xspace_{D_{\mu, K}}, g_{D_{\mu, K}}), R(\Xspace_{D_{\nu, K}}, g_{D_{\nu, K}}))\\
 		& \leq \|f-g\|_{\infty} + D_K(\mu, \nu), 
 		\end{align*}
 		where the last step uses~\cref{thm:stability_kernel_distance} and~\cref{lem:stability-local-smooth}.
 \end{proof}

 \subsection{Detailed Proofs for Section~\ref{sec:reeb-space}}
 \label{sec:reeb-space-proofs}
 In this section, we provide detailed proofs for the results in~\cref{sec:reeb-space}.


 \begin{proof}[Proof of Lemma~\ref{lem:reeb_space_smoothing}]
     We will prove this lemma through some results about quotient spaces (see \cite[Section~2.4]{engelking1989general}) and the fact that $X$ is compact and Hausdorff.
 	Consider the map $\Phi: \Xspace \times I_\varepsilon \to R(\Xspace, f) \times I_\varepsilon$ defined by $\Phi(x, t) = ([x], t)$ for any $(x, t)\in \Xspace \times I_\varepsilon$.
 	Then $\Phi$ is a continuous surjective map and function preserving with respect to $f_\varepsilon$ on $\Xspace \times I_\varepsilon$ and $f_\varepsilon$ on $R(\Xspace, f) \times I_\varepsilon$.
 	Since $\Phi$ preserves function values, so is the induced map $\tilde{\Phi}$.
 	Additionally, note that $\Phi$ induces a continuous surjective map $\tilde{\Phi}: R(\Xspace \times I_\varepsilon, f_\varepsilon) \to S_\varepsilon(R(\Xspace, f))$ defined by $\tilde{\Phi}([x, t]) = [[x], t]$.
 	Next, we show that $\tilde{\Phi}$ is a homeomorphism. This is done by showing that $\tilde{\Phi}$ is a bijective quotient map (\cite[Corollary~2.4.7]{engelking1989general}).
 	We will introduce some notations before showing that $\tilde{\Phi}$ is a quotient map.

 	We will use the notations $\pi_{f_\varepsilon}: \Xspace \times I_\varepsilon \to R(\Xspace \times I_\varepsilon, f_\varepsilon)$
 	and $\pi_{f_\varepsilon}: R(\Xspace, f) \times I_\varepsilon \to S_\varepsilon(R(\Xspace, f))$ to denote the quotient maps.
 	Then $\pi_{f_\varepsilon} \circ \Phi = \tilde{\Phi}\circ\pi_{{f}_\varepsilon}$.
 	 Note that as $\Phi$ is a continuous surjective map between compact Hausdorff spaces, $\Phi$ is closed and is a quotient map.
 	As the composition of two quotient maps, $\pi_{f_\varepsilon} \circ \Phi$ is also a quotient map.
 	Hence $\tilde{\Phi}$ is a quotient map by applying~\cite[Corollary~2.4.5]{engelking1989general} to the equality $\pi_{f_\varepsilon} \circ \Phi = \tilde{\Phi}\circ\pi_{{f}_\varepsilon}$.

 	As $\Phi$ is surjective so is $\tilde{\Phi}$.
 	Then it remains to show that $\tilde{\Phi}$ is injective.
 	That is for any $[x, t], [y, s] \in R(\Xspace \times I_\varepsilon, f_\varepsilon)$, if $\tilde{\Phi}([x, t]) = \tilde{\Phi}([y, s])$, then $[x, t] = [y, s]$.
 	Since $\tilde{\Phi}([x, t]) = \tilde{\Phi}([y, s])$, then we have $f_\varepsilon([x], t) = f_\varepsilon([y], s) = c$ for some $c\in \R$ and there exist a connected set $C\subset f_\varepsilon^{-1}(c)$ such that $[x, t], [y, s] \in C$.
 	Then $f(x) + t = f(y) + s$ which implies $f_\varepsilon(x, t) = f_\varepsilon(y, s)$.
 	As both $R(X\times I_\varepsilon, f_\varepsilon)$ and $S_\varepsilon(R(X, f))$ are compact Hausdorff spaces, the map $\tilde{\Phi}$ is closed.
 	Then $\tilde{\Phi}$ is a continuous closed quotient map such that for any point $[[x] , t]\in S_\varepsilon(R(\Xspace, f))$, the preimage $\tilde{\Phi}^{-1}([[x], t])$ is a connected set.
 	We then apply \cite[Theorem~2.4.4]{engelking1989general} to conclude that $\tilde{\Phi}^{-1}(C)$ is also connected.
 	Note that $\tilde{\Phi}^{-1}(C)$ belongs to $f_\varepsilon^{-1}(c)$ and contains both $[x, t]$ and $[y, s]$.
 	Then we have $[x, t] = [y, s]$.
 	This completes the proof.
 \end{proof}


 \begin{proof}[Proof of Theorem~\ref{thm:reeb_space_stability}]
 	Let $\varepsilon = \|f-g\|_\infty$.
 	We will prove the theorem by constructing the function preserving maps $\phi: \Xspace \to \Xspace \times I_{\varepsilon }$ and $\psi: \Xspace \to \Xspace \times I_{\varepsilon}$.
 	Specifically, we define $\phi$ and $\psi$ as follows:
 	\begin{equation*}
 		\phi(x) = (x, f(x) - g(x)) \quad \text{and} \quad \psi(x) = (x, g(x) - f(x)).
 	\end{equation*}
 	This is well-defined since $\|f(x) - g(x)\|_\infty \leq \|f-g\|_\infty$ for any $x\in \Xspace$.
 	Then $\phi$ is a function preserving map as $ g_\varepsilon\circ \phi(x) = g_\varepsilon(x, f(x) - g(x)) = g(x) + (f(x) - g(x)) = f(x)$. A similar calculation shows that $\psi$ is also a function preserving map. In particular, we will show that the following diagram commutes which in turn implies that $R(\Xspace, f)$ and $R(\Xspace, g)$ are $\varepsilon$-interleaved.
 	\[\begin{tikzcd}
 		{\Xspace} && {\Xspace\times I_\varepsilon} && {\Xspace\times I_{2\varepsilon}} \\
 		\\
 		{\Xspace} && {\Xspace\times I_\varepsilon} && {\Xspace\times I_{2\varepsilon}}
 		\arrow[from=1-1, to=1-3]
 		\arrow[from=1-3, to=1-5]
 		\arrow["{T_\varepsilon[{\phi}]}"'{pos=0.1}, color={rgb,255:red,92;green,92;blue,214}, dashed, from=1-3, to=3-5]
 		\arrow[from=3-3, to=3-5]
 		\arrow[from=3-1, to=3-3]
 		\arrow["{\phi}"'{pos=0.1}, color={rgb,255:red,92;green,92;blue,214}, from=1-1, to=3-3]
 		\arrow["{\psi}"{pos=0.1}, color={rgb,255:red,222;green,70;blue,23}, from=3-1, to=1-3]
 		\arrow["{T_\varepsilon[{\psi}]}"{pos=0.1}, color={rgb,255:red,222;green,70;blue,23}, dashed, from=3-3, to=1-5]
 		\arrow["{\eta}", from=1-1, to=1-3]
 		\arrow["{\eta_\epsilon}", from=1-3, to=1-5]
 		\arrow["{\eta}", from=3-1, to=3-3]
 		\arrow["{\eta_\epsilon}", from=3-3, to=3-5]
 	\end{tikzcd}\]
 	where $\eta$ and $\eta_\varepsilon$ are inclusion maps.
 	Indeed, we have
 	\begin{enumerate}
 		\item $T_\varepsilon[{\phi}]$ is a function preserving map,
 		\item $T_\varepsilon[{\psi}] \circ \phi = \eta_\epsilon \circ \eta$,
 		\item $T_\varepsilon[{\psi}]$ is a function preserving map,
 		\item $T_\varepsilon[{\phi}] \circ \psi = \eta_\epsilon \circ \eta$.
 	\end{enumerate}
 	By symmetry, we only need to show the first and the second items.
 	For the first item, we have
 	\begin{equation*}
 		\begin{aligned}
 			g_{2\varepsilon}\circ T_\varepsilon[{\phi}](x, t) & = g_{2\varepsilon}(x, f(x) - g(x) + t) \\
 			& = g(x) + (f(x) - g(x) + t) \\
 			& = f(x) + t.\\
 			& = f_\varepsilon(x, t).
 		\end{aligned}
 	\end{equation*}
 	For the second item, we have
 	\begin{equation*}
 		\begin{aligned}
 			T_\varepsilon[{\psi}] \circ \phi(x) & = T_\varepsilon[{\psi}](x, f(x) - g(x)) \\
 			& = (x, g(x) - f(x) + f(x) - g(x)) \\
 			& = (x, 0) \\
 			& = \eta_\varepsilon \circ \eta(x).
 		\end{aligned}
 	\end{equation*}
 	This completes the proof.
 \end{proof}

 \begin{proof}[Proof of Lemma~\ref{lem:stability-local-smooth-reeb-space}]
 	The proof is identical to the proof of Lemma~\ref{lem:stability-local-smooth} by simply viewing the parameter $t$ as an element in $\R^d$ instead of $\R$.
 \end{proof}

 \begin{proof}[Proof of Theorem~\ref{thm:stability-dtm-smoothed-reeb-space}]
 	The proof is identical to the proof of Theorem~\ref{thm:stability-dtm-smoothed} except that we need to use Lemma~\ref{lem:stability-local-smooth-reeb-space} instead of Lemma~\ref{lem:stability-local-smooth}.
 \end{proof}

 \begin{proof}[Proof of Theorem~\ref{thm:stability-dmk-smoothed-reeb-space}]
 	The proof is identical to the proof of Theorem~\ref{thm:stability-dmk-smoothed} except that we need to use Lemma~\ref{lem:stability-local-smooth-reeb-space} instead of Lemma~\ref{lem:stability-local-smooth}.
 \end{proof}

 \subsection{Detailed Proofs for Section~\ref{sec:range-integrated}}\label{sec:range-integrated-proofs}
 In this section, we provide detailed proofs for the results in~\cref{sec:range-integrated}.

 \begin{proof}[Proof of Proposition~\ref{prop:reeb-graph-range-measure-stability}]
 	By the triangle inequality of the interleaving distance and the stability of the Reeb graph (Theorem~\ref{thm:stability-reeb}), we have
     \begin{align*}
         d_I(R(X, F_\mu \circ f), R(X, F_\nu \circ g)) & \leq d_I(R(X, F_\mu \circ f), R(X, F_\mu \circ g)) \\
         & \quad\quad\quad+ d_I(R(X, F_\mu \circ g), R(X, F_\nu \circ g)) \\
         & \leq \|F_\mu \circ f - F_\mu \circ g\|_{\infty} + \|F_\mu \circ g - F_\nu \circ g\|_{\infty} \\
         & \leq \mathrm{Lip}(F_\mu)||f - g||_{\infty} + d_{KS}(\mu, \nu).
     \end{align*}
     Alternatively, there is a similar inequality
     \begin{align*}
         d_I(R(X, F_\mu \circ f), R(X, F_\nu \circ g)) & \leq d_I(R(X, F_\mu \circ f), R(X, F_\nu \circ f)) \\
         & \quad\quad\quad+ d_I(R(X, F_\nu \circ f), R(X, F_\nu \circ g)) \\
         & \leq \|F_\mu \circ f - F_\nu \circ f\|_{\infty} + \|F_\nu \circ f - F_\nu \circ g\|_{\infty} \\
         & \leq d_{KS}(\mu, \nu) + \mathrm{Lip}(F_\nu)||f - g||_{\infty}.
     \end{align*}
     Additionally, since the range of cumulative distribution function is $[0, 1]$, we have the simple bound $\|F_\mu \circ f - F_\nu \circ g \|_{\infty} \leq 1$.
     Therefore, we obtain the desired inequality by taking the minimum of the two bounds.
 \end{proof}

 \begin{proof}[Proof of Proposition~\ref{prop:reeb-graph-range-measure-critical-points}]
 	The regularity assumption on $\mu$ allows us to compute the critical points of $F_\mu \circ f$ through the chain rule.
     That is, the gradient $\nabla (F_\mu \circ f)$ is given by
     \begin{equation*}
         \nabla (F_\mu \circ f) = p(f) \nabla f,
     \end{equation*}
     where $p(f)$ is the density function of $\mu$ at $f$.
     As $p(f(x)) > 0$ for any $x \in X$, $\nabla (F_\mu \circ f)$ is zero if and only if $\nabla f$ is zero.
     Therefore, the critical points of $F_\mu \circ f$ are the same as the critical points of $f$.
     Additionally, for each critical point $x$ of $f$, the Hessian of $F_\mu \circ f$ at $x$ is given by
     \begin{equation*}
         \nabla^2 (F_\mu \circ f) (x) = p(f(x)) \nabla^2 f(x) + p'(f(x)) \nabla f(x) \otimes \nabla f(x) = p(f(x)) \nabla^2 f(x),
     \end{equation*}
     as $\nabla f(x) = 0$.
     Therefore, the Hessian of $F_\mu \circ f$ at critical points is non-degenerate as the Hessian of $f$ is non-degenerate and the number of positive and negative eigenvalues at each critical point is preserved.
     This implies that $F_\mu \circ f$ is Morse.
     Additionally, the critical values of $F_\mu \circ f$ are given by $F_\mu(f(x))$ where $x$ is a critical point of $f$.
 \end{proof}

 \subsection{Detailed Proofs for Section~\ref{sec:range-reeb-space}}
 \label{sec:range-reeb-space-proofs}
 In this section, we provide detailed proofs for the results in~\cref{sec:range-reeb-space}.

 \begin{proof}[Proof of Proposition~\ref{prop:reeb-space-range-measure-stability}]
     By the triangle inequality of the interleaving distance and the stability of the Reeb space (Theorem~\ref{thm:reeb_space_stability}), we have
     \begin{align*}
         d_I(R(X, F_\mu \circ f), R(X, F_\nu \circ g)) &\leq d_I(R(X, F_\mu \circ f), R(X, F_\mu \circ g))\\
         &\quad\quad\quad\quad\quad + d_I(R(X, F_\mu \circ g), R(X, F_\nu \circ g))\\
         &\leq \|F_\mu \circ f - F_\mu \circ g\|_\infty + \|F_\mu \circ g - F_\nu \circ g\|_\infty\\
         &\leq \mathrm{Lip}(F_\mu)||f - g||_{\infty} + \max_{1\leq i\leq d} |F_{\mu_i}\circ g - F_{\nu_i}\circ g|\\
         &\leq \mathrm{Lip}(F_\mu)||f - g||_{\infty} + \max_{1\leq i\leq d} \{
             d_{KS}(\mu_i, \nu_i)
         \}
     \end{align*}
     where the last inequality follows from the definition of the KS distance.
     Similarly, by utilizing the inequality $d_I(R(X, F_\mu \circ f), R(X, F_\nu \circ g)) \leq d_I(R(X, F_\mu \circ f), R(X, F_\nu \circ f)) + d_I(R(X, F_\nu \circ f), R(X, F_\nu \circ g))$, we can show that
     \begin{align*}
         d_I(R(X, F_\mu \circ f), R(X, F_\nu \circ g)) &\leq \mathrm{Lip}(F_\nu)||f - g||_{\infty} + \max_{1\leq i\leq d} \{
             d_{KS}(\mu_i, \nu_i)
         \}.
     \end{align*}
     Since the range of $F_\mu$ and $F_\nu$ are bounded in $[0, 1]^d$, we have $\|F_\mu\circ f - F_\nu\circ g\|_\infty \leq 1$.
     Then the proposition follows.
 \end{proof}

\end{document}